\documentclass[12pt]{elsarticle}
\usepackage{amsmath,amsfonts,amssymb,graphicx,hyperref,textcomp,color}

\journal{Commun Nonlinear Sci Numer Simulat}

\begin{document}

\begin{frontmatter}

\title{Adversarial decision strategies in multiple network phased oscillators: the Blue-Green-Red Kuramoto-Sakaguchi model}

\author[rvt]{Mathew Zuparic}
\ead{mathew.zuparic@dst.defence.gov.au}
\author[rvr]{Maia Angelova}
\ead{maia.a@deakin.edu.au}
\author[rvr]{Ye Zhu}
\ead{ye.zhu@deakin.edu.au}
\author[rvt]{Alexander Kalloniatis}
\ead{alex.kalloniatis@dst.defence.gov.au}
\address[rvt]{Defence Science and Technology Group, Canberra, ACT 2600, Australia}
\address[rvr]{School of Information Technology, Deakin University, Burwood, VIC 3125, Australia}

\begin{abstract}
We consider a model of three interacting sets of decision-making agents, labeled Blue, Green and Red, represented as 
coupled phased oscillators subject to frustrated
synchronisation dynamics. The agents are coupled on three networks of differing topologies,
with interactions modulated by different cross-population frustrations, internal and cross-network couplings.  
The intent of the dynamic model is to examine the degree to which 
two of the groups of decision-makers, Blue and Red, are able to realise a strategy of being ahead of each others' decision-making cycle while internally seeking synchronisation
of this process -- all in the context
of further interactions with the third population, Green.
To enable this analysis, we perform a significant dimensional reduction approximation
and stability analysis. We compare this to 
a numerical solution for a range of internal and cross-network coupling parameters to investigate various synchronisation regimes and critical thresholds. The comparison reveals good agreement for appropriate parameter ranges. Performing parameter sweeps, we reveal that Blue's pursuit of a strategy of staying too-far ahead of Red's decision cycles triggers a second-order effect of the Green population being ahead of Blue's cycles.
This behaviour has implications for the dynamics of
multiple interacting social groups with both cooperative and
competitive processes.
\end{abstract}


\end{frontmatter}


\section{Introduction}
\label{intro}

The spontaneous appearance of patterned behaviour in locally coupled dynamical systems is immensely relevant to social, biological, chemical and physical systems. Notable early examples of simple models that demonstrated such behaviours include Schelling's \cite{Schelling71} segregation models which display the emergence of communities based on the correlated choices and practices of individual decision-makers. Additionally, Watson and Lovelock's \textit{Daisyworld} model \cite{Watson83} demonstrates the growth and decline of different coloured flowers with different albedo levels that both compete for space and
coordinate to stabilise global temperatures simply by the flower's response to variable radiation levels received from the sun. 
Social versions of such patterned behaviour often
involve mixtures of competitive and cooperative dynamics.
Examples include the work of Abrams, Yaple and Wiener concerning 
religious affiliation \cite{Abrams2011}; 
social opinion dynamics using the Axelrod cultural model by Gonz\'{a}lez-Avella \textit{et al.}  \cite{Gonzalez-Avella14}; 
and examination of shifts in societal morals using networked Monte Carlo simulations and mean field theory by Vicente \textit{et al.} \cite{Vicente14}.
Common across all of these systems is the ability for seemingly \textit{unintelligent} actors as represented in 
components of dynamical or statistical physics models to display complex patterns and behaviours
within mathematical representations of aligned or
mis-aligned `intentions' or `strategies'. For a contemporary review of this topic of growing attention in the scientific community refer to Strogatz \cite{Strogatz03} and Chapter 2 of Ilachinski \cite{Ilachinski04}. In this paper, we extend
the approach of network synchronisation to modelling such complex systems with a dichotomy of cooperative and competitive processes across three sets of actors.

Using the \textit{Kuramoto model} \cite{Kuramoto84} as the starting point of this work, we focus on the onset of synchronisation amongst agent populations across multiple networks, where the agents exist in cooperative and adversarial relationships according to the degree of `frustration' in the interaction. The term frustration in this work is not used in an emotive sense, rather its application is similar to the term's use when applied to condensed matter systems, where atoms find themselves in non-trivial arrangements due to conflicting inter-atomic forces, usually referred to as geometrical frustration. 
Since its original inception, the Kuramoto model has provided a paradigmatic mathematical modelling environment to explore the onset of global critical phenomena; for recent reviews refer to \cite{Acebron05, Arenas08, Dorfler14, Rodrigues16, Dorogovtsev08}.
The role of frustration occurs in 
the Kuramoto-Sakaguchi model \cite{Sakaguchi86, Nicosia13, Kirkland15, Brede16, Coolen03}, where the introduction of phase shifts in the interaction terms changes
the potential steady-state behaviour from phase synchronisation (all phases equal) to frequency synchronisation (phases
shifted by a constant amount in relation to each other) between selected oscillators.
As this work is concerned with multiple populations, we focus on the \textit{multiple network} formulation of the model \cite{Kawamura10, Boccaletti14, Barreto08, Montbrio04, Lin09} where each sub-network has potentially different characteristics, such as graph topologies or natural frequency distributions. Notable examples of Kuramoto-based applications
to social-organisational systems can be found in; the conformists-and-contrarians
model \cite{HongStrogatz2011,Maistrenko2014}; the opinion-changing-rate model \cite{Pluchino06}; network community detection using the Fourier-Kuramoto model \cite{Maia17}; and the measurement and dynamic modelling of decision cycles in military headquarters \cite{SAWN17,Kalloniatis18a}.

In this work, we extend the two-network Blue-Red model of \cite{Kalloniatis16} to the three-network \textit{Blue-Green-Red} (BGR) model. The model's novelty comes from the introduction of the Green network, which is not on equal footing with Blue or Red; we impose that Green does not
`seek to be ahead of decisions' of either Blue or Red networks through a predefined strategy which
we characterise with the frustration parameter.
This is in contrast to the Blue-Red interaction,
as previously modelled in \cite{Kalloniatis16}. Nevertheless, as shall be shown in the following sections, Green still may stay ahead in phase
as a consequence of the nonlinear dynamics, but the mechanism for such a strategy comes from different sources. These mechanisms include other networks pursuing a certain strategy, and/or the structural choices Green makes with the way it interacts with Blue and Red. In each of these networks, we distinguish
`strategic' (or leadership) and `tactical' nodes. We also introduce an asymmetry into the model by imposing that the Blue and Green networks interact entirely through their strategic nodes, whereas Red and Green interact via their more numerous tactical nodes. This asymmetry allows analysis of the effect of exerting influence on senior decision-makers via the Blue-Green interaction, versus targeting the more numerous but less influential network members via the Green-Red interaction. A historical example includes the events during and after the 2001 Afghanistan war, where NATO/Coalition forces (Blue)
were engaging in military action against Taliban insurgents (Red) whilst concurrently
seeking to train wider Afghan society (Green) 
for their eventual assumption
of responsibility for the security
of their nation \cite{Allen2012}. Our interest in applying the Kuramoto model as a window into \textit{decision-making} processes is largely due to the cyclicity of the model's dynamic variables. While oscillations are pervasive in many physical, chemical
and biological systems \cite{Huard15}, the
human cognitive process also displays a fundamental cyclicity. Relevant versions of this process include the Perception Cycle model of Neisser \cite{Neisser76}, the Observe-Orient-Decide-Act (OODA) model of Boyd \cite{Osinga06}, and the Situation Awareness model of Endsley \cite{Endsley95}. For the majority of the paper, we analyse the model abstracted from the specific military application context, principally because the results have value for other applications of such a three-network model. 

A key result we find through both analytic and numerical examination is that there are regions of behaviour where Blue enjoys the advantage
over Red in being advance of the latter decision process.
However, within this, there are opportunities where
Green may be offered initiative by Blue, which
resonates with aspects of Counter-Insurgency strategy
\cite{Allen2012}. 

In the next section, we detail relevant parameters (networks, coupling, frequencies, frustrations) of the BGR model, and highlight how the asymmetry of the interaction of Green with both Blue and Red networks is manifested mathematically. We also detail a significant dimensional reduction technique which affords us semi-analytic insight into the dynamics. Section \ref{USECASE} provides the specific topologies of the networks, and input parameter choices for a use-case which runs throughout the remainder of the paper. In Section \ref{MODELAN} we provide a detailed analysis of the BGR model through the lens of specific network topologies and parameter choices. This includes comparing the semi-analytic outputs with the full numerical model, revealing very good agreement between both approaches, giving us the confidence to perform an extensive and computationally inexpensive parameter sweep of the model revealing areas of interest from each network's point of view. In the final Section we re-interpret the model
behaviours back in the context of the military application, and suggest future work.

\section{The Blue-Green-Red model}
\label{SEC2}
 
\subsection{Model definition}

The three-network BGR model is given by the following ordinary differential equations for each of the three sets of phases: Blue, Green and Red,  

\begin{eqnarray}
\dot{B}_i = \omega_i - \sigma_B \sum_{j \in {\cal B}}{\cal B}_{ij} \sin \left(B_i - B_j \right)- \zeta_{BG} \sum_{j \in {\cal G}}{\cal I}^{(BG)}_{ij} \sin \left(B_i - G_j - \phi_{BG} \right) \nonumber \\
- \zeta_{BR} \sum_{j \in {\cal R}}{\cal I}^{(BR)}_{ij} \sin \left(B_i - R_j - \phi_{BR} \right), \,\, i \in {\cal B},\label{MASTEREQ1}
\end{eqnarray}
\begin{eqnarray}
\dot{G}_i = \mu_i - \sigma_G \sum_{j \in {\cal G}}{\cal G}_{ij} \sin \left(G_i - G_j \right)- \zeta_{GB} \sum_{j \in {\cal B}}{\cal I}^{(GB)}_{ij} \sin \left(G_i - B_j  \right) \nonumber\\
- \zeta_{GR} \sum_{j \in {\cal R}}{\cal I}^{(GR)}_{ij} \sin \left(G_i - R_j  \right), \,\, i \in {\cal G},\label{MASTEREQ2}
\end{eqnarray}
\begin{eqnarray}
\dot{R}_i = \nu_i - \sigma_R \sum_{j \in {\cal R}}{\cal R}_{ij} \sin \left(R_i - R_j \right)- \zeta_{RB} \sum_{j \in {\cal B}}{\cal I}^{(RB)}_{ij} \sin \left(R_i - B_j - \phi_{RB} \right) \nonumber\\
- \zeta_{RG} \sum_{j \in {\cal G}}{\cal I}^{(RG)}_{ij} \sin \left(R_i - G_j - \phi_{RG} \right), \,\, i \in {\cal R},
\label{MASTEREQ3}
\end{eqnarray}
where each network's adjacency matrix is denoted by ${\cal B}, {\cal G}$ and ${\cal R}$. The dynamic variables $B_i, G_j$ and $R_k$ are the Blue, Green and Red phases,
or decision-states, for agents at each network's respective node $i \in {\cal B}$, $j \in {\cal G}$ and $k \in {\cal R}$. The variables $\omega_i$, $\mu_j$ and $\nu_k$ are the natural frequencies, or decision-speeds of 
the agents in isolation, with values typically drawn from a particular distribution. Furthermore, the parameters $\sigma_{B}$, $\sigma_G$ and $\sigma_R$ (all positive real valued) are referred to as the intra-network couplings, or intensity of interaction between agents. For one-network systems, the global coupling parameter controls the phase dynamics from a totally asynchronous regime to clustered limit cycles, and finally to phase locking behaviour \cite{Strogatz00, Restrepo05, Kalloniatis10, Taylor12, Dekker12}.

The inter-network adjacency matrices ${\cal I}^{(MN)}$ for networks ${\cal M}$ and ${\cal N}$ specify the connections between the nodes of network ${\cal M}$ and ${\cal N}$. Note that throughout this work we assume that ${\cal I}^{(MN)} = \left( {\cal I}^{(NM)} \right)^T$, though this assumption can be relaxed to offer more model generality. Furthermore, the inter-network couplings are specified by the parameters $\zeta_{MN} \in \mathbb{R}_+$, for networks ${\cal M}$ and ${\cal N}$. Lastly, the strategy chosen by agents of network ${\cal M}$ to collectively stay ahead of phase, or decision-state, of agents of network ${\cal N}$ is specified by the frustration parameter $\phi_{MN}\in \mathbb{S}^1$. We remark that the asymmetry between the Green network and Blue and Red is made clear in Eq.(\ref{MASTEREQ1}-\ref{MASTEREQ3}) by the absence of $\phi_{GB}$ and $\phi_{GR}$; this means that Green agents do not explicitly pursue a strategy to stay ahead in the phase of agents of other networks. We summarise the variables which comprise Eq.(\ref{MASTEREQ1}-\ref{MASTEREQ3}), and their interpretations, in Table \ref{tab:tab1}. 

\begin{table}[ht]
\caption{Summary of the variables used in Eq.(\ref{MASTEREQ1}-\ref{MASTEREQ3}), and their physical interpretations.} 
\centering 
\begin{tabular}{c c c } 
\hline 
expression & name & interpretation \\ [0.5ex] 
\hline 
$\{B,G,R\}$ & phase & dynamic agent decision state   \\
$\{{\cal B},{\cal G}, {\cal R}\}$ & adjacency matrix & internal network topology   \\
$\{\omega, \mu, \nu\}$ & natural frequency &  decision-speed of agent in isolation  \\
$\sigma_M$ & coupling of network ${\cal M}$ & intensity of agent interaction in ${\cal M}$\\
$ {\cal I}^{(MN)}$ & inter-adjacency matrix & topology between ${\cal M}$ and ${\cal N}$    \\
$ \zeta_{MN}$ & coupling between ${\cal M}$ and ${\cal N}$ & intensity of ${\cal M}$-${\cal N}$ agent interaction \\ 
$ \phi_{MN}$ & frustration & ${\cal M}$'s strategy against ${\cal N}$ \\ 
[1ex] 
\hline 
\end{tabular}
\label{tab:tab1} 
\end{table}

\begin{figure}
\begin{center}
\includegraphics[width=16cm]{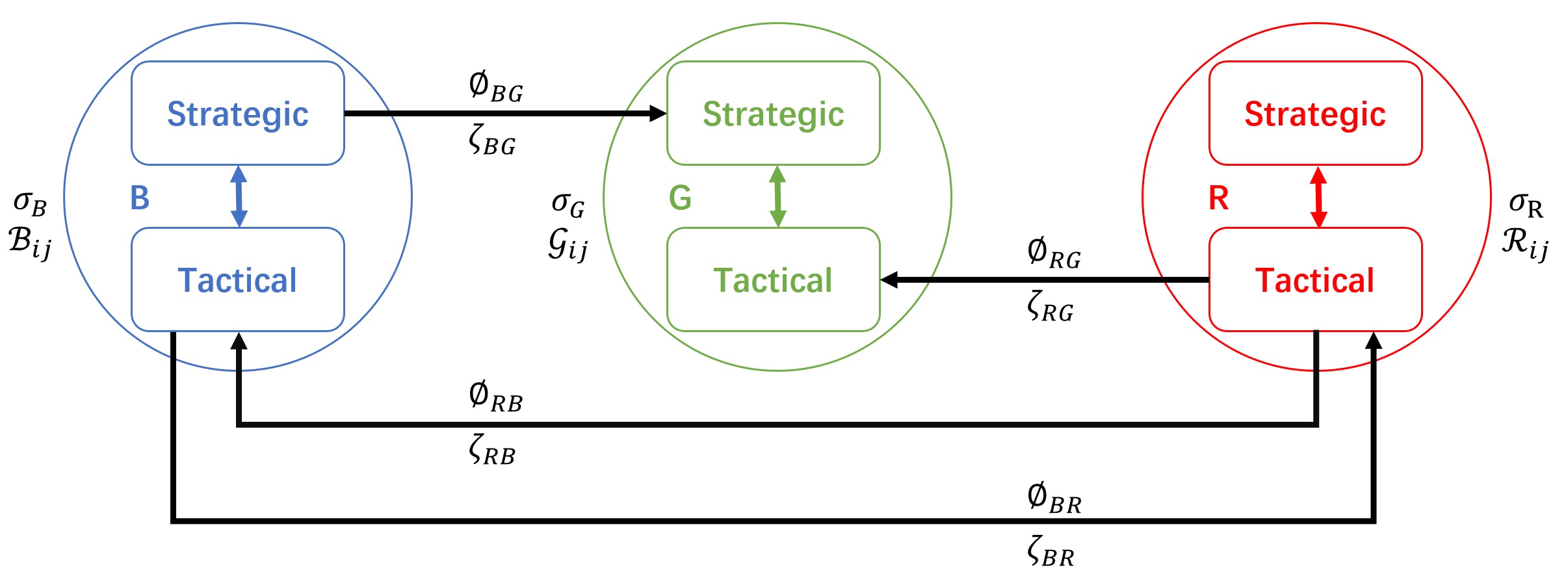}
\caption{Diagram of \textit{Blue}, \textit{Green} and \textit{Red}, their `strategic' and `tactical' substructures, and how they interact with one another.}
\label{fig:Model}
\end{center}
\end{figure}

A diagram of this scenario, with strategic and tactical sub-structures is shown in Figure \ref{fig:Model}. Strategic nodes for each network contain the highest number of connections in their respective graph, generally reflecting the span of control of leaders in social and organisational settings. The Blue and Red tactical networks interact with each other, attempting to stay ahead in the phase of their adversary's tactical nodes. In the absence of a Green network, the adversarial dynamics between Blue and Red networks has been explored in \cite{Kalloniatis16, Demazy18, Holder17, Kalloniatis18b}.

\subsection{Order parameters}
To measure the self-synchronisation within a given population, we use local order parameters for $\{B,G,R\}$ phases, labeled as $\{O_B, O_G, O_R\}$, respectively. The computation of the order parameters is accomplished using \textit{local} versions of Kuramoto's original \textit{global} order parameter \cite{Kuramoto84}:
\begin{equation}
    O_B = \frac{1}{|{\cal B}|} \left| \sum_{j \in {\cal B}} e^{i B_j} \right|, \;\; O_G = \frac{1}{|{\cal G}|} \left| \sum_{j \in G} e^{i G_j} \right|, \;\; O_R = \frac{1}{| {\cal R} |} \left| \sum_{j \in {\cal R}}  e^{i R_j} \right| .
\label{OrderINI}
\end{equation} 
The absolute value $|{\cal M}|$ represents the number of nodes of the generic graph ${\cal M}$. Values approaching unity
represents coherence of phases in the
respective networks, namely of synchronised
decision-making of the corresponding agents.

\subsection{Network centroids: semi-analytic approximation}
\label{SECT4}
By making the assumption that each of the three networks' phases has approximately synchronised, we can perform a significant dimensional reduction to Eq.(\ref{MASTEREQ1}-\ref{MASTEREQ3}). This is achieved by assuming the following form for each network's phase
\begin{equation}
B_i = b_i + E_B, \;\; G_j = g_j + E_G, \;\; R_k = r_k + E_R, \;\; \{i,j,k\} \in \{ \cal{B}, \cal{G}, \cal{R}\},
\label{approx}
\end{equation}
where $\{ b_i, g_j, r_k\}$  are `small' fluctuations,
namely $b^2_i \approx g^2_j \approx r^2_k \approx 0$. The variables $E_B, E_G$ and $E_R$ are the mean values of the phases for the Blue, Green and Red networks respectively,
\begin{equation}
E_B = \frac{1}{|\cal{B}|}\sum_{i \in {\cal B}} B_i,\;\; E_G = \frac{1}{|\cal{G}|}\sum_{j\in {\cal G}} G_j, \;\; E_R= \frac{1}{|\cal{R}|}\sum_{k\in {\cal R}} R_k.
\label{globalapp}
\end{equation}
$E_B$, $E_G$ and $E_R$ are alternatively referred to as the corresponding network's \textit{centroid}. The difference between each network's centroid value is denoted by
\begin{equation}
E_B - E_G \equiv \alpha_{BG}, \;\; E_G - E_R \equiv \alpha_{GR}, \;\; E_B - E_R \equiv \alpha_{BR} = \alpha_{BG} - \alpha_{GR}.
\label{angles}
\end{equation}
The approximations specified by Eq.(\ref{approx}) amount to a system of $|\cal{B}|+|\cal{G}|+|\cal{R}|$ defining equations, with $|{\cal B}|+|{\cal G}|+ |{\cal R}| +3$ variables. However, since $E_B$, $E_G$ and $E_R$ are the mean value of each network's phases, then necessarily we obtain that $\sum_{i \in {\cal B}} b_i = \sum_{j \in {\cal G}} g_j = \sum_{k \in {\cal R}}r_k = 0$, thus collapsing the system dimensionality appropriately. 

By inserting the approximation for the phases given by Eq.(\ref{approx}) into Eq.(\ref{MASTEREQ1}-\ref{MASTEREQ3}), and utilising properties of the eigenvalues and eigenvectors of the resulting graph-Laplacians (details shown in \ref{dimreduc}) we obtain the following expressions for the dynamics of the centroids
\begin{eqnarray}
\begin{split}
\dot{E}_B = \bar{\omega} - \frac{\zeta_{BG} d^{(BG)}_T}{|{\cal B}|} \sin \left(\alpha_{BG}-\phi_{BG} \right) - \frac{\zeta_{BR} d^{(BR)}_T}{|{\cal B}|} \sin \left( \alpha_{BR} - \phi_{BR} \right),\\
\dot{E}_G = \bar{\mu} + \frac{\zeta_{GB} d^{(GB)}_T}{|{\cal G}|} \sin  \alpha_{BG}  - \frac{\zeta_{GR} d^{(GR)}_T}{|{\cal G}|} \sin \alpha_{GR} ,\\
\dot{E}_R = \bar{\nu} + \frac{\zeta_{RB} d^{(RB)}_T}{|{\cal R}|} \sin ( \alpha_{BR} + \phi_{RB}) + \frac{\zeta_{RG} d^{(RG)}_T}{|{\cal R}|} \sin( \alpha_{GR} + \phi_{RG}),
\end{split}
\label{AllE}
\end{eqnarray}
where we have applied the notation
\begin{eqnarray}
\bar{\omega} \equiv \frac{1}{|{\cal B}|}\sum_{i \in {\cal B}} \omega_i,\;\; \bar{\mu} \equiv \frac{1}{|{\cal G}|}\sum_{i \in {\cal G}} \mu_i, \;\; \bar{\nu} \equiv \frac{1}{|{\cal R}|} \sum_{i \in {\cal R}} \nu_i,
\end{eqnarray}
for the respective means of each network's natural frequencies. Additionally,
\begin{equation}
d^{(MN)}_T \equiv \sum_{i \in {\cal M}} \sum_{k \in {\cal N}} {\cal I}^{(MN)}_{ik}, 
\end{equation}
is the total number of edges shared by networks ${\cal M}$ and ${\cal N}$. Eq.(\ref{AllE}) approximates the dynamics of the centroids of each of the three networks completely in terms of their differences. Taking the appropriate difference of each of the expressions in Eq.(\ref{AllE}) we collapse the dynamics of the centroids into the following two-dimensional system:
\begin{eqnarray}
\begin{split}
\dot{\alpha}_{BG} = \bar{\omega} - \bar{\mu} - \psi^B_G \sin(\alpha_{BG}-\phi_{BG}) - \psi^G_B  \sin \alpha_{BG}\\
- \psi^B_R \sin(\alpha_{BG}+\alpha_{GR}-\phi_{BR}) + \psi^G_R \sin\alpha_{GR} ,\\
\dot{\alpha}_{GR} = \bar{\mu} - \bar{\nu} +\psi^G_B\sin \alpha_{BG} -  \psi^R_B\sin(\alpha_{BG}+\alpha_{GR}+\phi_{RB})\\
-   \psi^G_R\sin \alpha_{GR} - \psi^R_G\sin(\alpha_{GR}+\phi_{RG}),
\end{split}\label{dynamic1}
\end{eqnarray}
where we have applied the notation,
\begin{equation}
\frac{d^{(M N )}_T\zeta_{M N}}{|\cal{M}|} = \psi^{M}_{N} \;\; \textrm{for networks}\;\; {\cal M} \;\; \textrm{and}\;\; {\cal N}.
\end{equation}
In Table \ref{tab:tab2} we offer a summary of the various measures which are applied in this work to analyse and understand model outputs. 

\begin{table}[ht]
\caption{Summary of the measures used to analyse model outputs.} 
\centering 
\begin{tabular}{c c c } 
\hline 
measure & name & range \\ [0.5ex] 
\hline 
$\{O_B,O_G,O_R\}$ & local order parameter & $(0,1)$   \\
$\{E_B,E_G,E_R\}$ & centroids/mean value of phases & $\mathbb{S}^1$   \\
$\{\alpha_{BG}, \alpha_{GR}, \alpha_{BR}\}$ & centroid differences & $\mathbb{S}^1$  \\
[1ex] 
\hline 
\end{tabular}
\label{tab:tab2} 
\end{table}

\section{Use-case}
\label{USECASE}
\subsection{Networks and natural frequencies}
\label{SECTnetandfreq}
\begin{figure}
\begin{center}
\includegraphics[width=16cm]{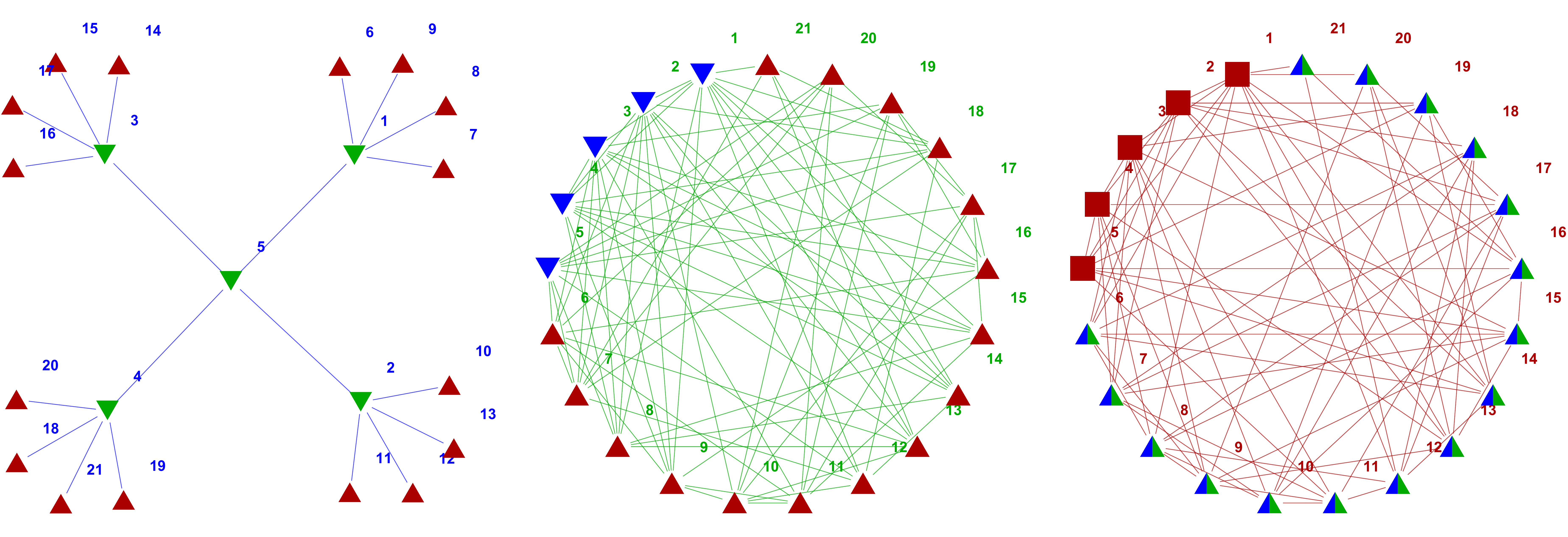}
\caption{Blue (hierarchy), Green (Watts-Strogatz) and Red (Erd\H{o}s-R\'{e}nyi) networks used in the numerical simulation of Eqs.(\ref{MASTEREQ1}--\ref{MASTEREQ3}). Nodes with the same label \textit{and} shape share edges with respective networks. For instance --- nodes 1--5 presented as upside-down triangles on the Blue network (coloured green) are linked with the correspondingly labeled nodes on the Green network, presented as blue upside-down triangles. Similarly --- nodes 6--21 presented as triangles on the Blue and Green networks (coloured red) are linked with the correspondingly labeled nodes on the Red network, presented as blue/green triangles. Nodes 1--5 on the Red network are the only nodes not externally connected with other networks.} 
\label{fig:Graphs}
\end{center}
\end{figure}

For numerical exploration of the BGR model, we construct graphs of size $|\cal{B}|=|\cal{G}|=|\cal{R}|$ = 21, given explicitly in Figure \ref{fig:Graphs}. This extends the example followed in previous Blue-vs-Red
studies in \cite{Kalloniatis16,Holder17,Kalloniatis18b}. 
As shown on the left side of Figure \ref{fig:Graphs}, the Blue population forms a \textit{hierarchy} stemming from a single root, followed by a series of four branches two layers deep. The right side of Figure \ref{fig:Graphs} shows the network for the Red population, given by a random \textit{Erd\H{o}s-R\'{e}nyi} graph, generated by placing a link between nodes with 0.4 probability. Finally, the network for the Green population, presented in the middle of Figure \ref{fig:Graphs}, is given by a small-world \textit{Watts-Strogatz} graph \cite{Newman00} with rewiring probability 0.3. These are all simplified caricatures
of, respectively, military, terrorist and societal structures
for the purpose of illustrating the
behaviours of the model.

Focusing on the Blue network on the left of Figure \ref{fig:Graphs}, the particular colour, shape and numbering of each node determines its connection to other graphs. Specifically, the nodes numbered 1--5 are coloured green, and hence each share an edge with the corresponding nodes on the network for Green which share the same number (1--5) and shape. Thus, the total number of connections between the Blue and Green networks is 5. Similarly, the red coloured triangle nodes, labeled 6--21, on both the Blue and Green networks are connected to the corresponding shaped and labeled nodes on the Red network, themselves coloured blue and green. Consequently, the total number of edges shared between the Blue-Red and Green-Red networks is 16. As indicated in Figure \ref{fig:Model}, the strategic nodes of Red, labeled 1--5 and portrayed as red squares, share no edges with either Blue or Green networks.

\begin{figure}
    \centering
    \includegraphics[width=16cm]{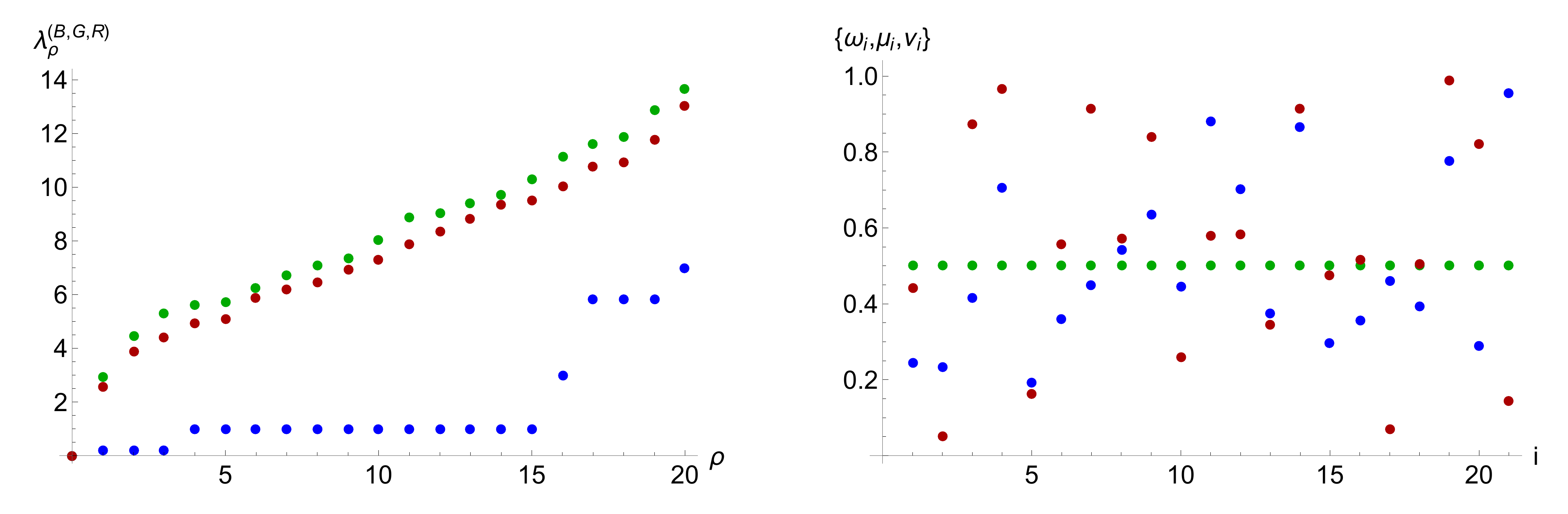}
    \caption{Left panel: The eigenvalue spectrum of the graph Laplacians (defined in Eq.(\ref{Laplace})) of the Blue, Green and Red networks. Right panel: The natural frequency values of the oscillators for each node on the Blue, Green and Red networks, with the mean-average of each network being $\bar{\omega} = 0.503$, $\bar{\mu}=0.500$ and $\bar{\nu}=0.551$ respectively.}
    \label{fig:evalandfreq}
\end{figure}

In the left panel of \ref{fig:evalandfreq} we present the eigenspectrum of the graph-Laplacians \cite{Bollabas98}, defined in Eq.(\ref{Laplace}), for the Blue, Green and Red networks, coloured accordingly. A key observation of the graph spectrum lies in the relatively lower eigenvalues of the Blue graph, which is a direct consequence of the poor connectivity afforded by a hierarchy (total number of edges equal to 20). Contrastingly, we see that the Green and Red networks possess very similar Laplacian eigenvalues, much higher than Blue, reflecting their relatively high connectivity, with a total number of edges of 84 and 77 for Green and Red respectively. 

The right panel of Figure \ref{fig:evalandfreq} gives the values of the natural frequencies used for each network's node. The frequency values for the Blue and Red networks were drawn from a uniform distribution between zero and unity, and for the Blue vs Red model \cite{Kalloniatis16, Holder17} the difference between the means of their respective frequencies, $\bar{\omega}-\bar{\nu}$, plays a critical role in the dynamics of the oscillators. Finally, for Green, the combination of the small-world topology, and the replicated natural frequencies for all the nodes, $\mu_i = 0.5$ $\forall \;i \in {\cal G}$, is chosen to emulate the Green network as a \textit{tight-knit community} \cite{Newman00}. Numerous works have shown that a well-connected network, with similar natural frequency values across the nodes, will have very good synchronisation properties. Thus, by placing Green in the middle of the adversarial relationship between Blue and Red, our intent is to examine the effect a tight-knit easily-synchronisable network has on the particular strategies chosen by the remaining adversarial networks.

\subsection{Coupling and frustration}
In order to make a meaningful comparison with previously published results \cite{Kalloniatis16, Holder17}, we apply the following intra-network coupling values:
\begin{equation}
    \sigma_B = 8, \;\; \sigma_G = 0.2, \;\; \sigma_R = 0.5,
\end{equation}
which are sufficient to enable the networks to internally synchronise without inter-network coupling. 
Observe here that the high coupling for Blue
compensates for the relatively poor connectivity of the hierarchy; this reflects the real-world phenomenon that
hierarchical organisations rely quite heavily on
tight discipline and training. Contrastingly, the lower
coupling of both Red and Green reflects the less disciplined
responsiveness between members of ad hoc organisations;
but their lower coupling is compensated by higher, if uneven connectivity. 
Additionally, we choose the inter-network coupling values:
\begin{equation}
    \zeta_{BR} = \zeta_{RB} = 0.4, \;\; \zeta_{BG} = \zeta_{GB}=\zeta_{GR}= \zeta_{RG} \equiv \zeta \in \mathbb{R},
    \label{couplechoice}
\end{equation}
The  main reason for these choices is that they are sufficiently high that
synchronisation is achievable, but also interesting deviations,
or disruptions to synchronisation may be detected and examined.
Furthermore, we choose the following values for the strategies of the adversarial networks:
\begin{equation}
    \phi_{BG} = \phi_{RB} = \phi_{RG} = 0, \;\; \phi_{BR} \in \mathbb{S}^1.
\end{equation}
To compare the outputs of Eq.(\ref{dynamic1}) with those of the full system given in Eq.(\ref{MASTEREQ1}-\ref{MASTEREQ3}) we set,
\begin{eqnarray}
\begin{split}
 d^{(BG)}_T=  d^{(GB)}_T = 5, \;\;d^{(BR)}_T=  d^{(RB)}_T = d^{(GR)}_T=  d^{(RG)}_T = 16,
\end{split}
\end{eqnarray}
which reflects the use-case topology explained in Figure \ref{fig:Graphs}. Thus the variables $\psi$ become
\begin{eqnarray}
\begin{split}
\psi^B_G =\psi^G_B \equiv \frac{5 }{21}\zeta,  \;\; \psi^B_R= \psi^R_B = \psi^R_G = \psi^G_R \equiv  \frac{16}{21} \zeta, \;\; \zeta \in \mathbb{R}
\end{split}
\end{eqnarray}
which allows us to understand model behaviour as we vary two key parameters: Blue's frustration with respect to Red $\phi_{BR}$, and the inter-network coupling $\zeta$. Thus Eq.(\ref{dynamic1}) becomes,
\begin{eqnarray}
\begin{split}
\dot{\alpha}_{BG}= 0.003  - \frac{10}{21}\zeta \sin \alpha_{BG} + \frac{16}{21}\zeta \left[ \sin \alpha_{GR} -  \sin (\alpha_{BG}+\alpha_{GR}- \phi_{BR}) \right] ,\\
\dot{\alpha}_{GR} =-0.051 +  \frac{5}{21}\zeta \sin \alpha_{BG} - \frac{16}{21} \zeta \left[2 \sin \alpha_{GR}  +  \sin (\alpha_{BG}+\alpha_{GR}) \right],
\end{split}\label{dynamic2}
\end{eqnarray}
which is easily solved numerically.

\section{Model analysis}
\label{MODELAN}
Code was developed in \textit{Matlab}\textsuperscript{\textregistered} 2017a, using the ODE 23tb package, to numerically solve Eq.(\ref{MASTEREQ1}-\ref{MASTEREQ3}) with initial conditions drawn from $(-\pi / 2, \pi /2)$. Critically, we were able to perform a simple validation of the code by reproducing the local order parameter trajectories for Blue and Red given in Figure 4 of \cite{Kalloniatis16} using the  parameter values $\zeta_{BG}=\zeta_{GR}=0$ and $\zeta_{BR}=\zeta_{RB}=0.4$, whilst varying $\phi_{BR}$. Notably, dynamic behaviour (limit cycles) from steady-state was detected for $\phi_{BR} > 0.950 \pi$ for these particular parameter values. 

For the full system of Eq.(\ref{MASTEREQ1}-\ref{MASTEREQ3}), outputs of Eq.(\ref{OrderINI}) while varying $\zeta \in (0,1]$ revealed that each of the networks had highly synchronised phase dynamics ($O \ge 0.95$) over this range. Although local phase synchronisation for each network is high, the centroids display dynamic limit-cycle behaviour for $\zeta \le 0.1$, and steady-state behaviour for $\zeta \ge 0.3$. The phase behaviour in the parameter range $0.1 < \zeta < 0.3$ is mixed, depending on the frustration value $\phi_{BR}$. Indeed, the behaviour amongst the centroids undergoes multiple transitions as the frustration parameter $\phi_{BR}$ varies. In order to explore the observation of multiple behavioural changes of the system, we use the approximation given by Eq.(\ref{dynamic2}) as the local synchronisation of each network is sufficiently high for the assumption given by Eq.(\ref{approx}) to hold. Example outputs of Eq.(\ref{OrderINI}) for the full system are given in \ref{APPC}.

\begin{figure}
\begin{center}
\includegraphics[width=16cm]{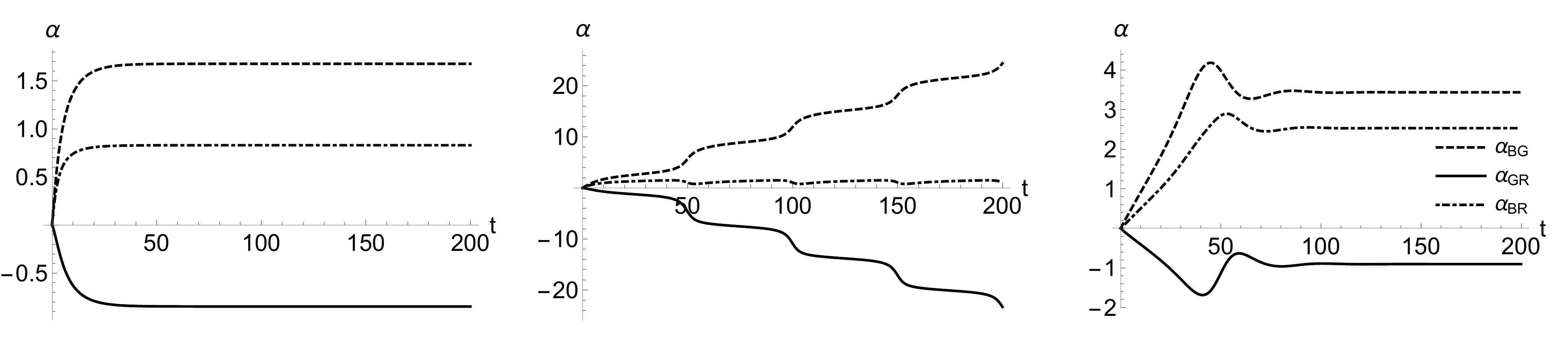}
\caption{Example of Eq.(\ref{dynamic2}), giving the difference of the centroids of the Blue, Green and Red networks for inter-network coupling value $\zeta =0.2$. Frustration parameter values from left-most to right-most columns are given by $\phi_{BR}= \{0.5 \pi, 0.7 \pi, 0.9 \pi\}$.} 
\label{fig:big3bodyrun}
\end{center}
\end{figure}
Figure \ref{fig:big3bodyrun} offers plots of the difference of the centroids given in Eq.(\ref{dynamic2}) for $\zeta = 0.2$, where frustration values are increased from left to right panels. The left-most panel ($\phi_{BR} = 0.5 \pi$) shows the three centroids in a steady-state (frequency-synchronised) arrangement with each other. Increasing $\phi_{BR}$ to $0.7 \pi$ in the middle panel, shows the system displaying limit cycle behaviour, with Green oscillating dynamically with respect to Blue and Red, who themselves have frequency synchronised with each other. Increasing $\phi_{BR}$ to $0.9 \pi$ in the right-most panel, the system returns to a steady-state regime. The three different modes of behaviour displayed while varying the frustration parameter suggest at least two values of $\phi_{BR}$ (for this particular value of $\zeta)$ which generate a regime change. We expose the mechanism of this regime change by careful examination of the steady-state solution(s) offered in Eq.(\ref{dynamic2}).

\subsection{Comparison between semi-analytic and fully numerical outputs}
\label{SECTcomparison}

\begin{figure}
\begin{center}
\includegraphics[width=16cm]{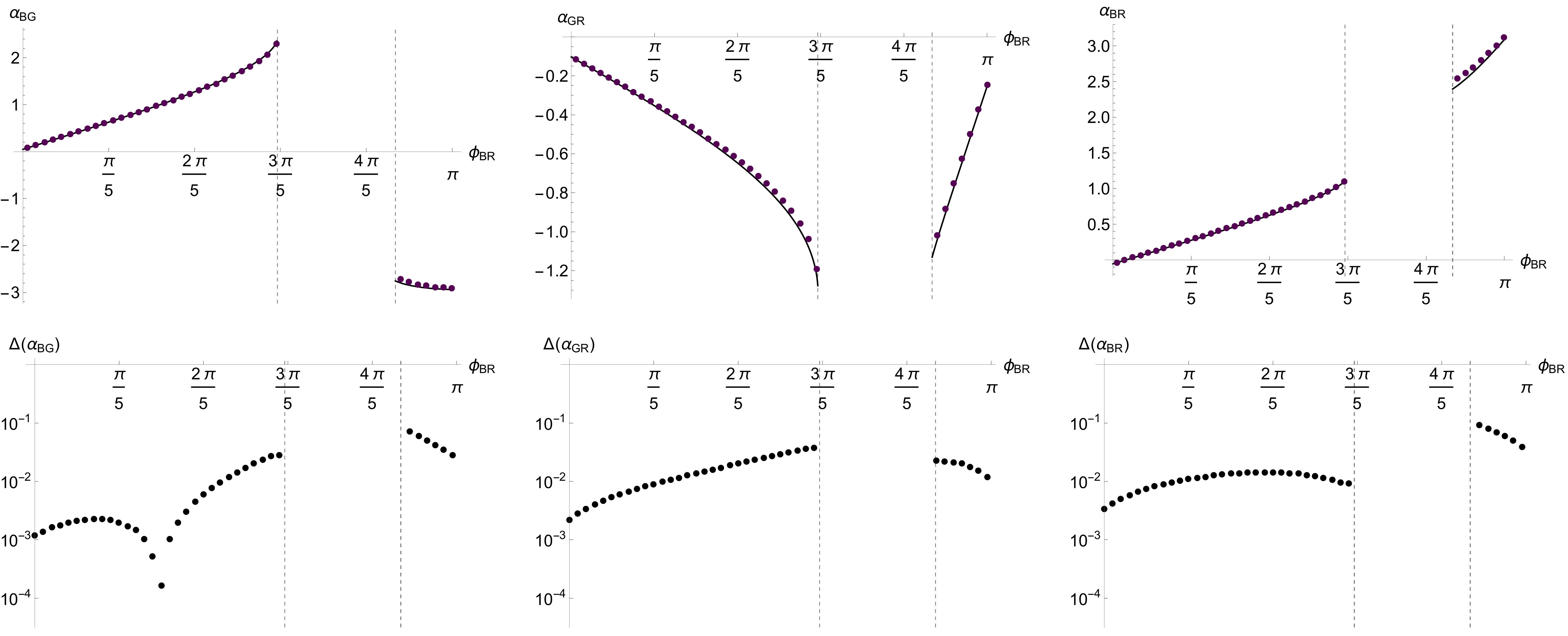}
\caption{Top row: plots showing the steady-state values
at $t=2000$ of the difference of the centroids for the three networks for $\zeta =0.2$ and varying $\phi_{BR} \in (0, \pi)$. Black lines show the semi-analytic solution resulting from Eq.(\ref{dynamic2}), and purple points give the equivalent fully numerical outcome. Note that there is no steady-state solution between $\phi_{BR} \in (0.593 \pi, 0.868 \pi)$ as the system in genuinely dynamic in that region. All solutions have been projected in the range $(-\pi,  \pi] \in \mathbb{S}^1$. Bottom row: logarithmic plots of the modulus of the difference between the semi-analytic and fully numerical outputs for the difference of the centroids of the networks --- labeled as $\Delta(\alpha)$.} 
\label{fig:alphasol02}
\end{center}
\end{figure}

Figure \ref{fig:alphasol02} offers a comparison between the two methods of solution, semi-analytic and fully numerical. The black line on the top row gives the steady-state position of $\alpha$ at $t=2000$ of the semi-analytic approach of Eq.(\ref{dynamic2}) for $\zeta =0.2$, whilst varying $\phi_{BR}$ as a continuous variable. Overlaid on these results appearing as purple points are the corresponding outputs from the fully numerical system. In order to account for any degeneracy introduced by the BGR model's trigonometric functions, the semi-analytic and fully numerical outputs are both projected onto $\mathbb{S}^1$ $(-\pi, \pi]$ via,
\begin{eqnarray}
 2 \arctan \left[ \tan \frac{\alpha_{BG}\left(\zeta,\phi_{BR}\right)}{2}\right]
\rightarrow 
\alpha_{BG} (\zeta, \phi_{BR}),
\label{projection}
\end{eqnarray}
and similarly for $\alpha_{GR}$ and $\alpha_{BR}$. The bottom row of Figure \ref{fig:alphasol02} presents the logarithmic plot of the modulus of the difference between the semi-analytic and fully numerical results for the difference of the centroids of the Blue, Green and Red networks, labeled as $\Delta(\alpha)$.

Focusing on the top row, the left-most panel of Figure \ref{fig:alphasol02} for $\alpha_{BG}$, displays an almost linear increase in the angle between the Blue and Green centroids as $\phi_{BR}$ increases in the range $(0,0.593 \pi)$. The system then enters a dynamic state for the parameter values $\phi_{BR} \in (0.593 \pi, 0.868\pi)$, represented in Figure \ref{fig:alphasol02} as gaps where no steady-state solution can be found. For the interval $\phi_{BR} \in (0.868 \pi, \pi)$, Eq.(\ref{dynamic2}) again enters a steady-state regime with $\alpha_{BR}$ being negative in $\mathbb{S}^1$ for this range of $\phi_{BR}$. Focusing on the bottom-left panel, the fully numerical results agree with the semi-analytic results when calculating $\alpha_{BG}$, with the largest divergence appearing immediately after the steady-state has been reestablished. The corresponding steady-state behaviour of $\alpha_{GR}$ and $\alpha_{BR}$ in the middle and right panels of Figure \ref{fig:alphasol02} similarly agrees with the semi-analytic computations. 

\subsection{Examination of the root system and stability}
\begin{figure}
\begin{center}
\includegraphics[width=14cm]{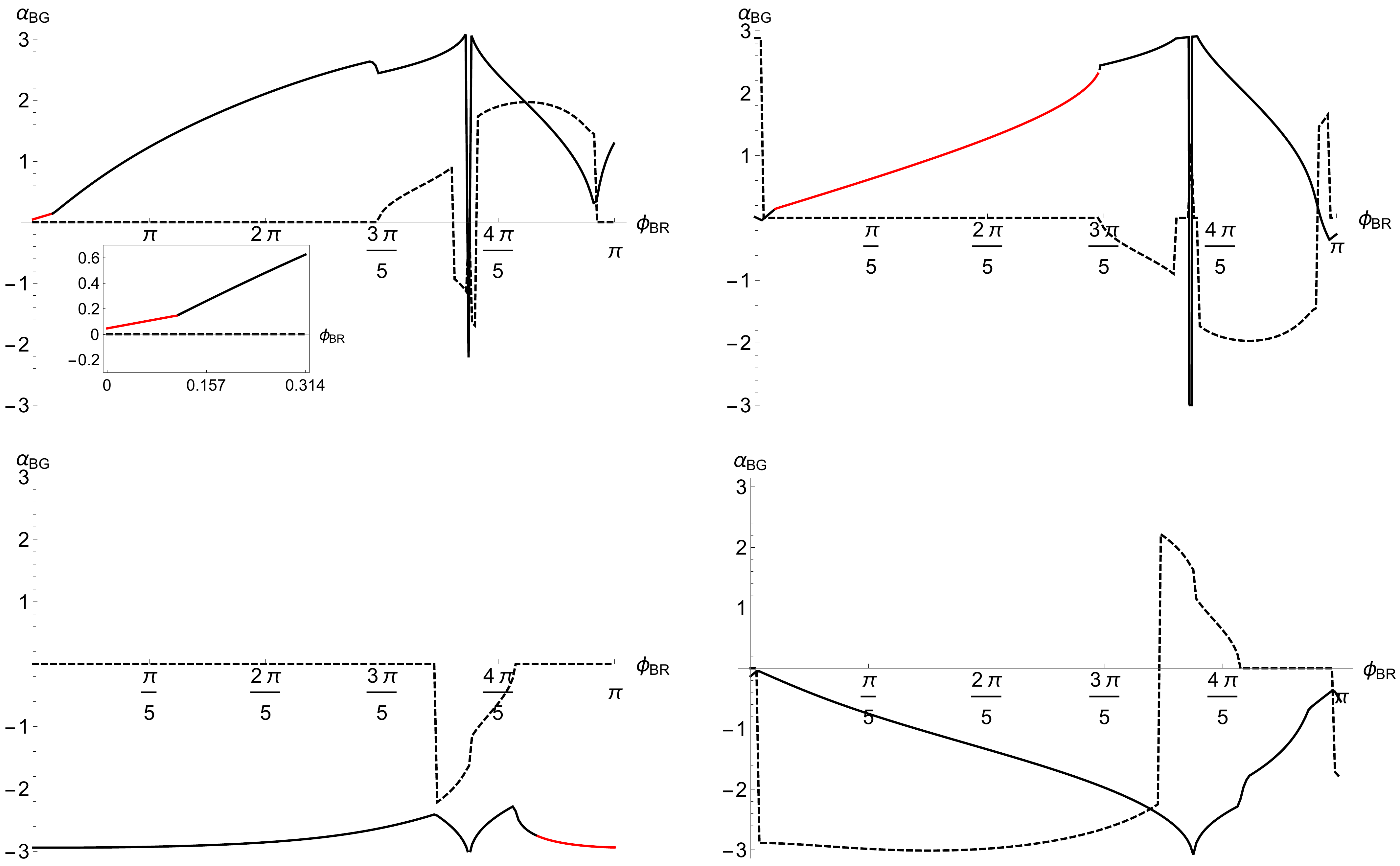}
\caption{Plots showing four of the six roots of $\alpha_{BG}$ in the system given in Eq.(\ref{static1}) for $\zeta =0.2$ and varying $\phi_{BR} \in [0,\pi]$. Solid and dashed curves denote the real and imaginary values of each of the roots, respectively. Note that the real sections given in red in the top row correspond to the steady-state solution given in the top-left panel in Figure \ref{fig:alphasol02}, with the inset in the top-left panel providing a zoomed-in perspective of the highlighted steady-state solution.} 
\label{fig:alphaBGroot}
\end{center}
\end{figure}

Figure \ref{fig:alphaBGroot} offers the fixed points of $\alpha_{BG}$ which arise
as {\it roots} of the following system, 
\begin{eqnarray}
\begin{split}
-\frac{0.063}{ \zeta} =   - 10 \sin \alpha_{BG} + 16 \left[ \sin \alpha_{GR} -  \sin (\alpha_{BG}+\alpha_{GR}- \phi_{BR})\right] ,\\
\frac{1.071 }{\zeta} = 5 \sin \alpha_{BG}- 16 \left[2 \sin \alpha_{GR}  +  \sin (\alpha_{BG}+\alpha_{GR})  \right],
\end{split}\label{static1}
\end{eqnarray}
for $\zeta=0.2$, and varying $\phi_{BR}$ continuously. Eq.(\ref{static1}) is obtained by inserting $\dot{\alpha}_{BG}=\dot{\alpha}_{GR} = 0$ in Eq.(\ref{dynamic2}). Furthermore, we project solutions for each of the roots onto $\mathbb{S}^1$ via Eq.(\ref{projection}). Figure \ref{fig:alphaBGroot} presents four of the six roots for $\alpha_{BG}$ that stem from  Eq.(\ref{static1}), containing both a real (solid curve) and imaginary (dashed curve) component for each root. The remaining two roots of $\alpha_{BG}$, and the six roots of $\alpha_{GR}$ (not shown) display qualitatively similar behaviour. The root values which coincide with the steady-state behaviour of Eq.(\ref{dynamic2}), given in the top-left panel of Figure \ref{fig:alphasol02}, are presented by the red line
sections in the top-left, top-right and bottom-left panels of Figures \ref{fig:alphaBGroot}.

\begin{figure}
\begin{center}
\includegraphics[width=15cm]{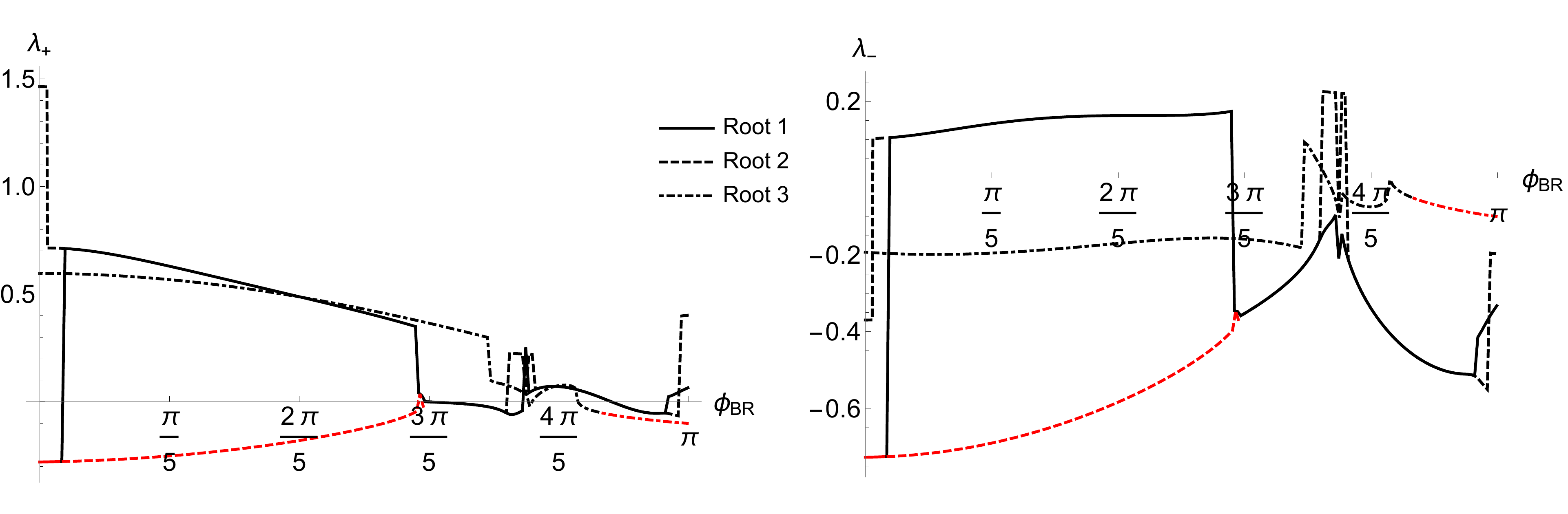}
\caption{Plots of the real components of the Lyapunov exponents detailed in the linearised system given in Eq.(\ref{lyapunov}) for $\zeta =0.2$ and varying $\phi_{BR} \in [0,\pi]$. Each of the three lines correspond to the three different roots which have valid sections for different values of $\phi_{BR}$. Similar to Figure \ref{fig:alphaBGroot}, the red sections of each of the three curves detail the exponents which are valid across the $\phi_{BR}$ region.} 
\label{fig:lyapunov}
\end{center}
\end{figure}

The reason for the solution jumping from one root to another is not immediately comprehensible from these plots. To this end, we perform stability analysis by substituting $\alpha_{BG}= \alpha^*_{BG}+\delta_1$ and $\alpha_{GR}= \alpha^*_{GR}+\delta_2$ into Eq.(\ref{dynamic2}), where the constant terms $\alpha^*_{BG}$ and $\alpha^*_{GR}$ are the roots of the system (shown in Figure \ref{fig:alphaBGroot} for $\alpha_{BG}$). We also assume that the time-dependent perturbations $\delta_1$ and $\delta_2$ are small, \textit{i.e.} $\delta^2_1 \approx \delta_1 \delta_2 \approx \delta^2_2 \approx 0$. Thus, Eq.(\ref{dynamic2}) becomes,
\begin{eqnarray}
\begin{split}
\dot{\delta}_1=  \bar{\omega}-\bar{\mu}-2 \psi^B_G \sin \alpha^*_{BG} +\psi^G_R \sin \alpha^*_{GR}  - \psi^B_R \sin\left( \alpha^*_{BR} -\phi_{BR}  \right) +\beta_{11} \delta_1 + \beta_{21} \delta_2,\\
\dot{\delta}_2=  \bar{\mu}-\bar{\nu} + \psi^B_G \sin \alpha^*_{BG} -2 \psi^G_R \sin \alpha^*_{GR}  - \psi^B_R \sin \alpha^*_{BR} +\beta_{21} \delta_1 + \beta_{22}\delta_2,
\end{split}
\label{stab1}
\end{eqnarray}
where
\begin{eqnarray}
\begin{split}
\beta_{11} = -\left[ 2 \psi^B_G \cos \alpha^*_{BG} +\psi^B_R \cos \left( \alpha^*_{BR}-\phi_{BR} \right) \right],\\
\beta_{12} = \psi^G_R \cos \alpha^*_{GR} -  \psi^B_R \cos \left(  \alpha^*_{GR} -\phi_{BR}\right),\\
\beta_{21} = \psi^B_G \cos \alpha^*_{BG} -  \psi^B_R \cos  \alpha^*_{GR},\\
\beta_{22} = -\left( 2 \psi^G_R \cos \alpha^*_{GR} +\psi^B_R \cos \alpha^*_{BR} \right).
\end{split}
\label{stab2}
\end{eqnarray}
Hence the Lyapunov exponents of the linearised system are
\begin{equation}
    \lambda_{\pm} = \frac{\beta_{11}+\beta_{22}}{2} \pm \sqrt{\frac{\left( \beta_{11}-\beta_{22} \right)^2}{4} + \beta_{12} \beta_{21} }.
    \label{lyapunov}
\end{equation}
The Lyapunov exponents which corresponds to each of the valid roots as a function of $\phi_{BR}$ are given in Figure \ref{fig:lyapunov}. Valid root values corresponding to the correct steady-state solution must satisfy:
\begin{itemize}
\item{zero imaginary component of the root values, and;}
\item{negative real values of the Lyapunov exponents $\lambda_+$ and $\lambda_-$.}
\end{itemize}
Given these requirements it is possible to choose the correct roots in the region $\phi_{BR} \in (0,0.593 \pi) \cup (0.868 \pi, \pi)$, due to there being only one root which fulfils all the requirements. It is also possible to determine that there are no valid roots in the region $\phi_{BR} \in (0.593 \pi, 0.830 \pi)$, as in this region none of the roots satisfies the stability requirements. 
Nevertheless, the linearised system detailed in Eq.(\ref{stab1}) and (\ref{stab2}) is not sensitive enough to detect limit cycles in the region $\phi_{BR} \in (0.830 \pi, 0.868 \pi)$; a small discrepancy
is visible in the right-most red section of Figure \ref{fig:lyapunov}, with its onset after the change of sign. Indeed, all of the requirements are satisfied in this region (real-valued roots and negative real components of the Lyapunov exponents), yet we know from Figure \ref{fig:alphasol02} that this region displays limit cycle behaviour. We also tested the sensitivity of the stability analysis by adding additional terms to Eq.(\ref{stab1}). The addition of quadratic terms did not increase Eq.(\ref{stab1})'s ability to detect limit cycles in this region, whereas with the addition of cubic terms we were only able to additionally detect limit cycle behaviour in the region $\phi_{BR} \in (0.830 \pi, 0.834 \pi)$. We forego these details for the sake of brevity.


\subsection{Contour plots}
\begin{figure}
\begin{center}
\includegraphics[width=16cm]{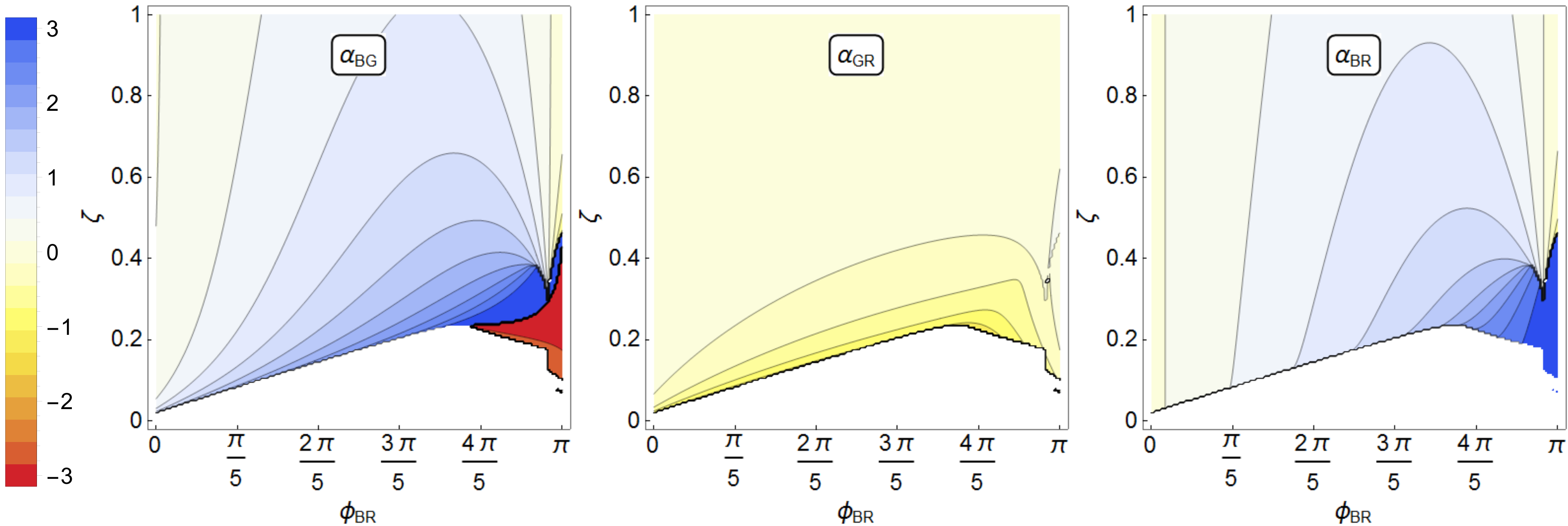}
\caption{Contour plots of Eq.(\ref{dynamic2}) for $\alpha_{BG}$, $\alpha_{GR}$ and $\alpha_{BR}$, varying both $\zeta \in [0,1]$ and $\phi_{BR}\in [0,\pi]$. Each panel is generated by calculating the $t=2000$ values of $\alpha_{BG}$ and $\alpha_{GR}$ in Eq.(\ref{dynamic2}), varying values of $\zeta$ and $\phi_{BR}$ over an equally spaced $201$ by $201$ grid. Points categorised as dynamic are shown as white. Points categorised as steady state are shown in colour and projected onto $\mathbb{S}^1$ via Eq.(\ref{projection}).} 
\label{fig:contourplot}
\end{center}
\end{figure}
We conclude this section by presenting contour plots, given by Figure \ref{fig:contourplot}, of Eq.(\ref{dynamic2}) for $\alpha_{BG}$, $\alpha_{GR}$ and $\alpha_{BR}$, varying both $\zeta \in [0,1]$ and $\phi_{BR}\in [0,\pi]$ as continuous variables. Each panel in Figure \ref{fig:contourplot} was generated by calculating the $t=2000$ values of $\alpha_{BG}$ and $\alpha_{GR}$ in Eq.(\ref{dynamic2}), varying values of $\zeta$ and $\phi_{BR}$ over an equally spaced $201$ by $201$ grid, and projected onto $\mathbb{S}^1$ using Eq.(\ref{projection}). Each point on the contour plot was suitably tested whether it could be categorised as either steady-state or dynamic. Dynamic points appear as white in Figure \ref{fig:contourplot}, whereas steady-state values are coloured based on the legend appearing on the very left of Figure \ref{fig:contourplot}. 


In the left panel of Figure \ref{fig:contourplot}, for values of inter-network coupling $\zeta \in [0.6,1]$, $\alpha_{BG}$ is mostly positive, rarely rising above a value of unity, except for a small region where $\phi_{BR} \approx \pi$, where the value of $\alpha_{BG}$ is negative, but still small. In this region of $\zeta$-values, one of the effects of Blue intending to be approximately $\pi$ ahead of Red is for Green to suddenly be ahead of Blue, as indicated by the appearance of colours corresponding to negative values. Additionally, we note that in this region of $\zeta$ values, the contours for $\alpha_{BG}$ vary quite smoothly. Contours start displaying more varied behaviour as the value of $\zeta$ decreases, with contour lines becoming denser. Generally, as $\zeta$ values decrease, we witness both greater rates of change as $\zeta$ and $\phi_{BR}$ vary, and the appearance of more extreme values of $\alpha_{BG}$. In the region $\zeta \in [0.2,0.4]$ and $\phi_{BR} \in [0.8 \pi, \pi]$, the value of $\alpha_{BG}$ varies quite drastically across \textit{all} values between $(-\pi, \pi)$. Of course, below a certain threshold of $\zeta$, which is influenced by the value of $\phi_{BR}$, the system becomes dynamic. Intuitively, we note that between $\phi_{BR} \in [0, 0.8 \pi]$, a higher value $\phi_{BR}$ requires a higher value of $\zeta$ to enable a steady-state solution to Eq.(\ref{dynamic2}); demonstrating that a greater frustration value generally places more strain on the system, which then requires greater coupling strength to enable steady-state solutions. For $\phi_{BR} > 0.8 \pi$, highly negative steady-state solutions appear for lower $\zeta$ values showing that the system has flipped with
Blue behind Green.

The middle panel of Figure \ref{fig:contourplot}, showing $\alpha_{GR}$, demonstrates less dramatic behaviour than that witnessed in the left panel. Generally, as $\zeta$ decreases below $0.6$, $\alpha_{GR}$ becomes more negative, very rarely reaching values less than $-1$. There is a small region however where $\alpha_{GR}$ exhibits small positive values for $\zeta \in [0.4 \pi, 0.6 \pi]$ and $\phi_{BR} \approx \pi$. Generally, however, rates of contour changes as $\zeta$ and $\phi_{BR}$ vary in the middle panel are never as extreme as witnessed in regions of the left panel for $\alpha_{BG}$. 

The right panel for $\alpha_{BR}$ is given by the addition of both left and middle panels. Visually, the right panel is very similar to the left panel for $\alpha_{BG}$, but lacks the more extreme rates of change as $\zeta$ and $\phi_{BR}$ vary. In the small region $\zeta \in [0.2,0.4]$ and $\phi_{BR} \in [0.8 \pi, \pi]$, the value of $\alpha_{BR}$ reveals Blue's phase to be maximally ahead of that of Red. Nevertheless, in the same parameter region for $\alpha_{BG}$, we see that the \textit{exact opposite} can occur: $\alpha_{BG}$ shows Blue centroid to be minimised in relation to Green. This phenomenon is an interesting second-order effect caused by the Blue network's frustration in relation to the Red network in a model where
the interactions of a third population are intrinsic.

\section{Conclusions, discussion and future work}
We have extended the adversarial two-network Blue-Red model of locally coupled frustrated phase oscillators to include a third networked population of actors
with vanishing frustration. Through numerical analysis and dimensional reduction we found that as frustrations increase or inter-population couplings decrease, the system discontinuously
flips, where the steady-state phase advance of one population 
in relation to another would increase, become time-varying, and then reverse.
Notably, Green's behaviour showed that in certain parameter ranges 
it may be ahead of Blue's centroid, despite vanishing frustration parameter.
The `sweet spot', where both Blue could be ahead in the
phase of Red and Green, and Green in turn ahead of Red,
was very narrow.


We can interpret these results, particularly 
the contour plot Fig.\ref{fig:contourplot}, 
through the lens that
frustration represents a strategy for decision advantage
and the BGR model captures multiple adversarial/cooperative decision-making groups.
Firstly, it shows that
even an organisational structure such
as a hierarchy --- which is 
designed for equal distribution of
information loads and a single source
of intent, but intrinsically
poorly structured for
synchronised decision-making
against
better-connected adversaries ---
can achieve advantageous outcomes.
However, to achieve this it requires
tighter internal and external
interaction.
There are significant portions of
parameter space in
Fig.\ref{fig:contourplot} with high $\zeta$
where Blue is both ahead of Green's
decision-making (thus able to exert
influence on a neutral party) and
an adversary. It is intuitively
plausible that as Blue seeks to be further
ahead of decision of both Green and Red
(through greater $\phi$) then
it must also increase its intensity
of interaction $\zeta$ to maintain
a steady-state decision advantage.

Blue may find itself behind
Green, even though it maintains a decision advantage with respect to 
Red, as seen in the extreme right
region of the first panel of Fig.\ref{fig:contourplot}. We remark that this is where Blue gains its most extreme advantage over Red in the third panel. Interpreted in the context of multi-party decision-making, this offers an interesting option
for Blue: that sometimes
ceding initiative to the neutral group provides scope for a maximal advantage over a competitor.
This phenomenon offers a qualitative (and,
to the degree that parameters in the model may 
eventually be measured
in real-world systems, quantitative) means to examine the risks introduced by pursuing a particularly greedy strategy: by striving to be too far ahead of a competitor's decision making, a population may allow non-adversarial actors to be ahead of one's decision-making processes. Like all risks, however, this phenomena can also be seen as an opportunity; does Blue use the fact that Green is afforded a means to stay ahead in decision-making cycles as a method of gaining trust with the third population? An answer in the affirmative or negative is of course context-sensitive, depending on the application. 

The model offers an intuitive conclusion: that in contexts of multiple parties with non-consistent
objectives, the sweet spot of unilateral advantage
for one party over the others may be very narrow
or non-existent altogether. 
The implications of such decision positions cannot
be deduced from within the model itself but how
it is embedded in the real world context --- either
through qualitative considerations, or by coupling
these behaviours to an additional mathematical model of
the external actions in the world.

Significantly, we do not observe in the regimes of
semi-stable behaviour in any regions where Red is ahead of Blue,
even though for the two-network case such behaviours can
be found \cite{Kalloniatis16}. We have not performed complete
parameter sweeps of the BGR model, so such regions
may exist. Alternately, the introduction of
the third population in the particular {\it asymmetric} way implemented here 
may push such behaviours into a narrow island inside more
chaotic dynamics. This is worth further numerical investigation
but may be outside the scope for an analytical solution.
However, this does imply the value of
strategic engagement with neutral parties in such a three-way
contested context.

Future work may consider stochastic noise in the BGR model as a means to explore the effects of uncertainty of human decision making in an adversarial engagement. Furthermore, it may be meaningful to frame the BGR model in a game-theory setting; the utility functions of such a study, and their measurement, may yield novel and useful ways to think about risk and trust between noncombatant groups caught up in inherently adversarial settings. Finally, the coupling of this model
into a representation of the outcomes of decisions
will yield a means of quantifying risks through
the interplay between probability and consequences.
In particular, in view of the military
contextualisation we adopt with this
model there is an opportunity
to couple this model with well-known
mathematical representations of combat
and network generalisations of them
\cite{Kalloniatis2020}.
Above all, through a compact mathematical model
of complexity
such as this, at least partially analytical
insights may be gained into
otherwise surprising and rich behaviours.

\section*{Acknowledgements}
The authors would like to thank Richard Taylor, Irena Ali and Hossein Seif Zadeh for discussions during the writing of this manuscript. This research was a collaboration between the Commonwealth of Australia (represented by the Defence Science and Technology Group) and Deakin University through a Defence Science Partnerships agreement. 

\appendix
\section{Strategic-tactical view} \label{Appendix1}
Following Figure \ref{fig:Model}, namely the segregation of strategic and tactical nodes, we offer the BGR model as the following \textit{expanded} set of ordinary differential equations, segregated into the relevant strategic (labelled by $I$) and tactical (labelled by
$II$) components,
\begin{eqnarray}
\begin{split}
\dot{B}^{(I)}_i = \omega^{(I)}_i - \sigma_B \sum_{j \in {\cal B}^{(I)}}{\cal B}_{ij} \sin \left(B^{(I)}_i - B^{(I)}_j \right) -\sigma_B \sum_{j \in {\cal B}^{(II)}}{\cal B}_{ij} \sin \left(B^{(I)}_i - B^{(II)}_j \right)\\
- \zeta_{BG} \sum_{j \in {\cal G}^{(I)}}{\cal I}^{(BG)}_{ij} \sin \left(B^{(I)}_i - G^{(I)}_j - \phi_{BG} \right), \,\, i \in {\cal B}^{(I)},\\
\dot{B}^{(II)}_i = \omega^{(II)}_i - \sigma_B \sum_{j \in {\cal B}^{(I)}}{\cal B}_{ij} \sin \left(B^{(II)}_i - B^{(I)}_j \right) -\sigma_B \sum_{j \in {\cal B}^{(II)}}{\cal B}_{ij} \sin \left(B^{(II)}_i - B^{(II)}_j \right)\\
- \zeta_{BR} \sum_{j \in {\cal R}^{(II)}}{\cal I}^{(BR)}_{ij} \sin \left(B^{(II)}_i - R^{(II)}_j - \phi_{BR} \right), \,\, i \in {\cal B}^{(II)},
\end{split}
\label{BLUE}
\end{eqnarray}
\begin{eqnarray}
\begin{split}
\dot{G}^{(I)}_i = \mu^{(I)}_i - \sigma_G \sum_{j \in {\cal G}^{(I)}}{\cal G}_{ij} \sin \left(G^{(I)}_i - G^{(I)}_j \right) -\sigma_G \sum_{j \in {\cal G}^{(II)}}{\cal G}_{ij} \sin \left(G^{(I)}_i - G^{(II)}_j \right)\\
- \zeta_{GB} \sum_{j \in {\cal B}^{(I)}}{\cal I}^{(GB)}_{ij} \sin \left(G^{(I)}_i - B^{(I)}_j \right), \,\, i \in {\cal G}^{(I)},\\
\dot{G}^{(II)}_i = \mu^{(II)}_i - \sigma_G \sum_{j \in {\cal G}^{(I)}}{\cal G}_{ij} \sin \left(G^{(II)}_i - G^{(I)}_j \right) -\sigma_G \sum_{j \in {\cal G}^{(II)}}{\cal G}_{ij} \sin \left(G^{(II)}_i - G^{(II)}_j \right)\\
- \zeta_{GR} \sum_{j \in {\cal R}^{(II)}}{\cal I}^{(GR)}_{ij} \sin \left(G^{(II)}_i - R^{(II)}_j \right), \,\, i \in {\cal G}^{(II)},
\end{split}
\label{GREEN}
\end{eqnarray}
\begin{eqnarray}
\begin{split}
\dot{R}^{(I)}_i = \nu^{(I)}_i - \sigma_R \sum_{j \in {\cal R}^{(I)}}{\cal R}_{ij} \sin \left(R^{(I)}_i - R^{(I)}_j \right) -\sigma_R \sum_{j \in {\cal R}^{(II)}}{\cal R}_{ij} \sin \left(R^{(I)}_i - R^{(II)}_j \right), \,\, i \in {\cal R}^{(I)}, \\
\dot{R}^{(II)}_i = \nu^{(II)}_i - \sigma_R \sum_{j \in {\cal R}^{(I)}}{\cal R}_{ij} \sin \left(R^{(II)}_i - R^{(I)}_j \right) -\sigma_B \sum_{j \in {\cal R}^{(II)}}{\cal R}_{ij} \sin \left(R^{(II)}_i - R^{(II)}_j \right)\\
- \zeta_{RB} \sum_{j \in {\cal B}^{(II)}}{\cal I}^{(RB)}_{ij} \sin \left(R^{(II)}_i - B^{(II)}_j - \phi_{RB} \right)\\
- \zeta_{RG} \sum_{j \in {\cal G}^{(II)}}{\cal I}^{(RG)}_{ij} \sin \left(R^{(II)}_i - G^{(II)}_j - \phi_{RG} \right), \,\, i \in {\cal R}^{(II)}.
\end{split}
\label{RED}
\end{eqnarray}
Eqs.(\ref{BLUE}-\ref{RED}) explicitly highlights the roles and interactions of the types of nodes in the networks by designating the strategic and tactical nodes.

\section{Dimensional reduction}
\label{dimreduc}
Inserting the approximation detailed in Eq.(\ref{approx}) into Eq.(\ref{MASTEREQ1}-\ref{MASTEREQ3}) we obtain,
\begin{eqnarray}
\begin{split}
\dot{E}_B + \dot{b}_i = \omega_i - \sigma_B \sum_{j \in \cal{B}} L^{(B)}_{ij}b_j - \zeta_{BG}\sin (\alpha_{BG} - \phi_{BG}) d^{(BG)}_i - \zeta_{BR}\sin (\alpha_{BR} - \phi_{BR}) d^{(BR)}_i\\
-\zeta_{BG} \cos (\alpha_{BG} - \phi_{BG}) \sum_{j \in {\cal B} \cup {\cal G} } {\cal L}^{(BG)}_{ij} V_j -\zeta_{BR} \cos (\alpha_{BR} - \phi_{BR}) \sum_{j \in {\cal B} \cup {\cal R} } {\cal L}^{(BR)}_{ij} V_j,\\
\dot{E}_G + \dot{g}_i = \mu_i - \sigma_G \sum_{j \in \cal{G}} L^{(G)}_{ij}g_j + \zeta_{GB}\sin \alpha_{BG}  d^{(GB)}_i - \zeta_{GR}\sin \alpha_{GR}  d^{(GR)}_i\\
+\zeta_{GB} \cos \alpha_{BG}  \sum_{j \in {\cal B} \cup {\cal G}} {\cal L}^{(GB)}_{ij} V_j -\zeta_{GR} \cos \alpha_{GR}  \sum_{j \in {\cal G} \cup {\cal R} } {\cal L}^{(GR)}_{ij} V_j,\\
\dot{E}_R + \dot{r}_i = \nu_i - \sigma_R \sum_{j \in \cal{R}} L^{(R)}_{ij}r_j + \zeta_{RB}\sin (\alpha_{BR} + \phi_{RB}) d^{(RB)}_i + \zeta_{RG}\sin (\alpha_{GR} + \phi_{RG}) d^{(RG)}_i\\
+\zeta_{RB} \cos (\alpha_{BR} + \phi_{RB}) \sum_{j \in {\cal B} \cup {\cal R}} {\cal L}^{(RB)}_{ij} V_j + \zeta_{RG} \cos (\alpha_{GR} + \phi_{RG}) \sum_{j \in {\cal G} \cup {\cal R}} {\cal L}^{(RG)}_{ij} V_j,
\end{split}
\label{BigApprox}
\end{eqnarray}
where $\{L^{(B)}, L^{(G)}, L^{(R)} \}$ are the graph Laplacians \cite{Bollabas98} of the Blue, Green and Red networks respectively:
\begin{equation}
    L^{(B)}_{ij} = \underbrace{\sum_{k \in {\cal B}} {\cal B}_{ik}}_{\equiv d^{(B)}_i} \delta_{ij} - {\cal B}_{ij},\;\; L^{(G)}_{ij} = \underbrace{\sum_{k \in {\cal G}} {\cal G}_{ik}}_{\equiv d^{(G)}_i} \delta_{ij} - {\cal G}_{ij},\;\; L^{(R)}_{ij}= \underbrace{\sum_{k\in {\cal R}} {\cal R}_{ik}}_{\equiv d^{(R)}_i} \delta_{ij} - {\cal R}_{ij}.
    \label{Laplace}
\end{equation}
Correspondingly, the matrices ${\cal L}^{(BG)}, {\cal L}^{(GB)}$ \textit{etc}. are the inter-network graph Laplacians, given by,
\begin{equation}
    {\cal L}^{(BG)}_{ij} = \underbrace{\sum_{k \in {\cal B} \cup {\cal G}} {\cal I}^{(BG)}_{ik} }_{=d^{(BG)}_i}  \delta_{ij}- {\cal I}^{(BG)}_{ij} , \;\; {\cal L}^{(GB)}_{ij} = \underbrace{ \sum_{k \in {\cal B} \cup {\cal G} } {\cal I}^{(GB)}_{ik}}_{=d^{(GB)}_i} \delta_{ij}   - {\cal I}^{(GB)}_{ij}.
\end{equation}
and similarly for $(BR), (GR), (RB)$ and $(RG)$. The integer $d_i$, for node $i$, is the \textit{degree} of node $i$ (total number of edges) for the particular network or inter-network connection. Lastly, the quantity $V_i$ in Eq.(\ref{BigApprox}) simply encodes the fluctuations for each network,
\begin{equation}
    V_i = \left\{\begin{array}{cc}
    b_i & i \in {\cal B}\\
    g_i & i \in {\cal G}\\
    r_i & i \in {\cal R}
    \end{array}.\right.
\end{equation}
The intra-network Laplacians present in Eq.(\ref{BigApprox}) all come equipped with a complete spanning set of orthonormal eigenvectors, which we label by
\begin{eqnarray}
\begin{split}
e^{(B,\rho_1)}_i, &\;\; \rho_1 = 0,1,\dots,|{\cal B}|-1 \in {\cal B}_E, & \sum_{j \in {\cal B}} L^{(B)}_{ij} e^{(B,\rho_1)}_j = \lambda^{(B)}_{\rho_1} e^{(B,\rho_1)}_i, \\
e^{(G,\rho_2)}_j, &\;\; \rho_2 = 0,1,\dots,|{\cal G}|-1 \in {\cal G}_E, & \sum_{j \in {\cal G}} L^{(G)}_{ij} e^{(G,\rho_2)}_j = \lambda^{(G)}_{\rho_2} e^{(G,\rho_2)}_i, \\
e^{(R,\rho_3)}_k, &\;\; \rho_3 = 0,1,\dots,|{\cal R}|-1 \in {\cal R}_E, & \sum_{j \in {\cal R}} L^{(R)}_{ij} e^{(R,\rho_3)}_j = \lambda^{(R)}_{\rho_3} e^{(R,\rho_3)}_i,
\end{split}
\end{eqnarray}
where we distinguish between indices in the eigen-mode space $\{ {\cal B}_E, {\cal G}_E, {\cal R}_E \}$ and those in the node space $\{ {\cal B}, {\cal G}, {\cal R} \}$. The spectrum of Laplacian eigenvalues of any given network, labeled $\{ \lambda^{(B)}, \lambda^{(G)}, \lambda^{(R)}\}$, is real-valued and conveniently bounded from below by zero; the degeneracy of the zero eigenvector equals the number of components of the respective network \cite{Bollabas98}. Thus, the Blue, Green and Red networks given in Figure \ref{fig:Graphs} each contain a single zero-valued eigenvalue --- for the Laplacian eigenvalues of the particular networks used in this work refer to the left panel of Figure \ref{fig:evalandfreq}. The corresponding zero eigenvectors $\{e^{(B,0)}, e^{(G,0)}, e^{(R,0)}\}$, up to normalisation, consist entirely of unit valued entries.

We wish to use the completeness of the Laplacians to
diagonalise the system. For a single network, namely the ordinary
Kuramoto-Sakaguchi model for a single graph, the Laplacian basis 
elegantly separates out the collective mode, corresponding to the synchronised
system, which identifies with the Laplacian zero eigenvector. Contrastingly,
the non-zero, or `normal', modes turn out to be Lyapunov stable, namely exponentially suppressed.
Thus the Laplacian neatly exposes the dynamics close to synchrony for the ordinary Kuramoto model
\cite{Kalloniatis10}.

In the case of multiple networks, the Laplacians do not commute and
therefore do not provide a simultaneous diagonalisation of the system.
To proceed with the dimensional reduction procedure we impose the further approximation
that 
\begin{eqnarray}
\sum_{j \in {\cal M} \cup {\cal N}}{\cal L}^{(MN)}_{ij} V_j \approx 0, \;\; \forall \;\; \textrm{networks} \;\; {\cal M} \;\; \textrm{and} \;\; {\cal N},
\label{harshapprox}
\end{eqnarray}
i.e. all inter-network Laplacian fluctuations in Eq.(\ref{BigApprox}) are approximately equal to zero. This approximation enables the fluctuations $b_i, g_i$ and $r_i$ in Eq.(\ref{BigApprox}) to decouple. Nevertheless, as mentioned in \cite{Holder17}, this approximation is not guaranteed to completely hold, even in model regimes which enable Eq.(\ref{approx}) to be satisfied. 

We now expand the fluctuations in Eq.(\ref{BigApprox}) via the non-zero normal modes,
\begin{eqnarray}
b_i = \sum_{\rho \in {\cal B}_E / \{0\}} e^{(B,\rho)}_i x_{\rho}, \;\; g_i = \sum_{\rho \in {\cal G}_E / \{0\}} e^{(G,\rho)}_i y_{\rho}, \;\; r_i = \sum_{\rho \in {\cal R}_E / \{ 0 \}} e^{(R,\rho)}_i z_{\rho},
\end{eqnarray}
and exploit the orthonormality of the spanning vectors to obtain,
\begin{eqnarray}
\begin{split}
    \dot{x}_{\rho} = q^{(\rho)}_B(x_{\rho}, \alpha_{BG},\alpha_{GR}) \;\; \dot{y}_{\rho} = q^{(\rho)}_G (y_{\rho}, \alpha_{BG},\alpha_{GR}) , \;\; \dot{z}_{\rho} =  q^{(\rho)}_R(z_{\rho}, \alpha_{BG}, \alpha_{GR}),
    \label{linear0}
\end{split}
\end{eqnarray}
where,
\begin{eqnarray}
\begin{split}
    q^{(\rho)}_B = \sum_{i \in {\cal B} }  e^{(B,\rho)}_i \left[\omega_i -\zeta_{BG} \sin ( \alpha_{BG} - \phi_{BG}) d^{(BG)}_i - \zeta_{BR} \sin(\alpha_{BR} - \phi_{BR}) d^{(BR)}_i \right]\\
    - \sigma_B \lambda^{(B)}_{\rho} x_{\rho},\;\; \rho \in {\cal B}_E / \{0\},\\
    q^{(\rho)}_G = \sum_{i \in {\cal G}}  e^{(G,\rho)}_i \left[ \mu_i + \zeta_{GB} \sin \alpha_{BG}  d^{(GB)}_i - \zeta_{GR} \sin \alpha_{GR}  d^{(GR)}_i \right] - \sigma_G \lambda^{(G)}_{\rho} y_{\rho}, \;\; \rho \in {\cal G}_E / \{0\}, \\
    q^{(\rho)}_R = \sum_{i \in {\cal R}}  e^{(R,\rho)}_i \left[ \nu_i + \zeta_{RB}\sin (\alpha_{BR} + \phi_{RB})d^{(RB)}_i - \zeta_{RG} \sin ( \alpha_{GR} + \phi_{RG}) d^{(RG)}_i  \right] \\
    - \sigma_R \lambda^{(R)}_{\rho} z_{\rho}, \;\; \rho \in {\cal R}_E / \{0\}.
\end{split}
\label{linear1}
\end{eqnarray}
We note that the time derivative of the centroids is eliminated from each expression in Eq.(\ref{linear1}) due to the orthogonality of the Laplacian eigenvectors; the sum of the individual entries of each non-zero Laplacian eigenvector exactly equals zero. Eqs.(\ref{linear0},\ref{linear1}) give
the dynamics of the normal modes of the BGR system. Eqs.(\ref{linear0},\ref{linear1}) are linear in their respective fluctuation mode variables, but ultimately their dynamics involves the differences of the centroids $\alpha_{BG}$ and $\alpha_{GR}$. These variables themselves are completely determined by the two-dimensional system in Eq.(\ref{dynamic1}), which is a two-dimensional extension of a \textsl{tilted periodic ratchet} system \cite{Reimann02, Zuparic17}.  

Finally,  projecting Eq.(\ref{BigApprox}) 
onto the zero eigenvectors for each of Blue, Green and Red we obtain the expressions for the centroids given in Eq.(\ref{AllE}) in the main text. Because the zero eigenvector projection separates
out equations for $E_B$, $B_G$ and $E_R$ in Eq.(\ref{AllE}),
we may refer to these as the zero-mode projections of the phases $B$, $G$ and $R$, respectively.

\section{Local synchronisation examples}
\label{APPC}
\begin{figure}[ht]
    \centering
    \includegraphics[width=16cm]{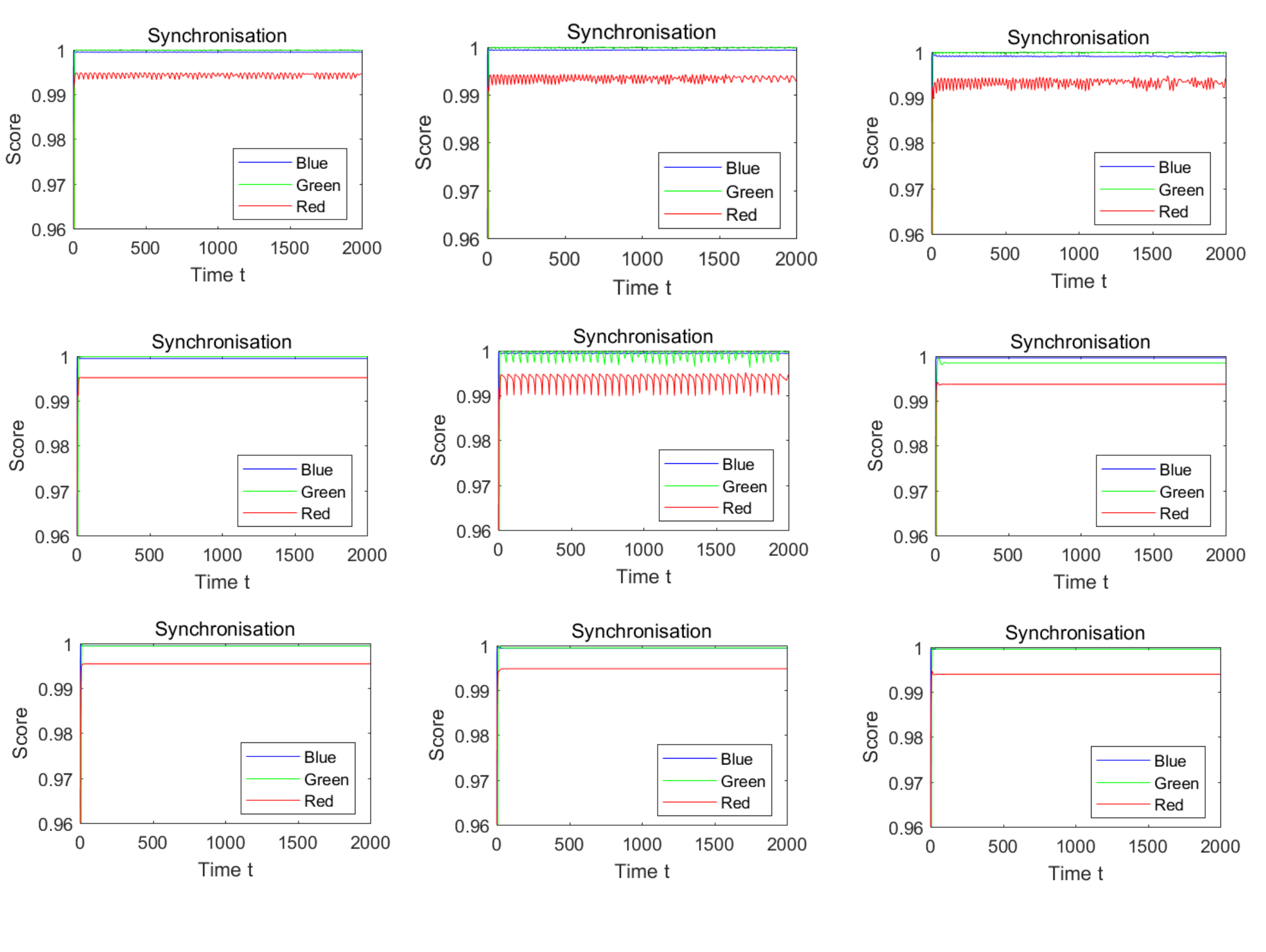}
    \caption{Example of Eq.(\ref{OrderINI}), giving the local synchronisation order parameter outputs for the Blue, Green and Red networks. Inter-network coupling values from top-most to bottom-most rows are given by $\zeta=\{0.1,0.2,0.3 \}$. Frustration parameter values from left-most to right-most columns are given by $\phi_{BR}=\{0.5 \pi, 0.7\pi ,0.9 \pi \}$.}
    \label{fig:mainnumer}
\end{figure}

Figure \ref{fig:mainnumer} offers numerical outputs of local order parameter values of all three networks. In the top row, for $\zeta = 0.1$ all three networks display high-frequency limit cycle behaviour while highly synchronised internally. The effect of increasing the frustration parameter $\phi_{BR}$ from the left-most to the right-most panel on the top row does not appear to have an appreciable influence on this behaviour other than making it
slightly more erratic. The second row of Figure \ref{fig:mainnumer}, for $\zeta = 0.2$, presents a more interesting picture. The local order parameters switch between steady-state behaviour for the left-most panel ($\phi_{BR}=0.5\pi$), to periodic limit cycle behaviour on the middle panel ($\phi_{BR}=0.7\pi$), and back to steady-state behaviour for the right-most panel ($\phi_{BR}=0.9\pi$). The bottom row of Figure \ref{fig:mainnumer}, for $\zeta = 0.3$, produces steady-state model outputs, regardless of the value of the Blue network's strategy towards Red.

\end{document}